\documentclass[12pt]{article}
\usepackage{epsfig}
\usepackage{psfrag}
\newcommand{\be}{\begin{eqnarray}}
\newcommand{\ba}{\begin{array}}
\newcommand{\ea}{\end{array}}
\newcommand{\ee}{\end{eqnarray}}

\newcommand{\beq}{\begin{equation}}
\newcommand{\eeq}{\end{equation}}
\newcommand{\beqa}{\begin{eqnarray}}
\newcommand{\eeqa}{\end{eqnarray}}

\newcommand{\no}{\nonumber}

\newcommand{\vs}{\vspace{-0.25cm}}

\begin{document}

\begin{flushright}
{\small  FZJ-IKP(TH)-1999-06}
\end{flushright}

\vspace{0.4in}

\begin{center}

{{\Large\bf 
Pion--nucleon scattering inside\\[0.3em]
the Mandelstam triangle}}\footnote{Work 
supported in part by DFG under contract no. ME 864-15/1.}

\end{center}

\vspace{.3in}

\begin{center}
{\large 
Paul B\"uttiker\footnote{email: P.Buettiker@fz-juelich.de},
Ulf-G. Mei{\ss}ner\footnote{email: Ulf-G.Meissner@fz-juelich.de}
}

\bigskip

{\em Forschungszentrum J\"ulich, Institut f\"ur Kernphysik 
(Theorie)\\ D-52425 J\"ulich, Germany}

\end{center}

\vspace{.9in}

\thispagestyle{empty} 

\begin{abstract}\noindent
We study the third order pion--nucleon scattering amplitude
obtained from heavy baryon chiral perturbation theory
inside the Mandelstam triangle. We reconstruct the pion--nucleon
amplitude in the unphysical region by use of dispersion relations
and determine the pertinent low--energy constants by a fit to
this amplitude. A detailed comparison with values obtained from
phase shift analysis is given. Our analysis leads to a pion--nucleon
$\sigma$--term of $\sigma (0) = 40\,$MeV based on the Karlsruhe
partial wave analysis. We have repeated the same procedure using the
latest solution of the VPI group and find a much larger value for
$\sigma (0)$.
\end{abstract}

\vfill

\pagebreak

\section{Introduction}
\label{sec:intro}

\begin{figure}
\begin{center}
\epsfig{file=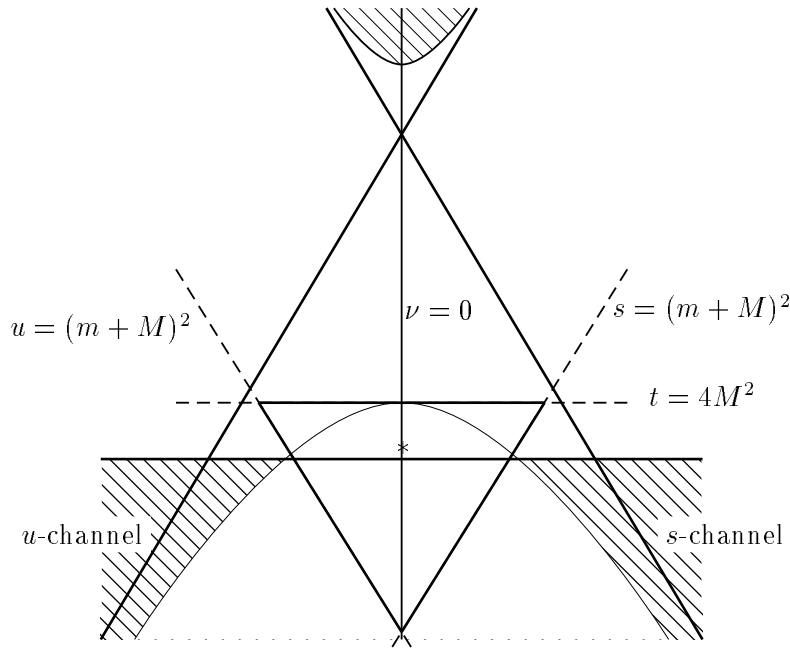}
\caption{Mandelstam plane. The Mandelstam triangle is the inside of the
thick lines. The star marks the so--called ``ideal point'' explained in 
sec.~\ref{sec:fits}.}
\label{fig:mandel}
\end{center}
\end{figure}

A detailed understanding of elastic pion--nucleon scattering
in the low energy region allows for precise tests of the chiral QCD dynamics.
Recently, this process has been investigated
to third order in heavy baryon chiral perturbation theory by various groups.
At that order, one has to deal with tree graphs and one loop diagrams.
At second and third order, there are in total four and five so--called low--energy
constants, respectively, which have to be determined by comparison to
data. The loop contribution
only depends on the weak pion decay constant $F=92.4$~MeV and the 
axial--vector coupling
$g_A = 1.26$. In ref.\cite{BKMpin}, particular combinations of threshold and 
subthreshold
parameters were found which do not depend on the dimension three low--energy
constants (LECs). Also, it was shown that the ensuing numerical values of these 
dimension two LECs, called $c_i$, can be understood in terms of resonance
exchange. In ref.\cite{moj}, the 
threshold parameters based on the Karlsruhe--Helsinki analysis together with the
pion--nucleon $\sigma$--term were used to fix the nine LECs. The results for the
$c_i$ were in good agreement with the ones obtained in \cite{BKMpin}. The amplitude
given in \cite{moj} was  used later to obtain S-- and P--wave phase shifts below
the first resonance~\cite{dapa}. The most systematic study was performed in 
ref.\cite{FMS}, where the S-- and P--wave partial wave amplitudes from three
different analyses were used (in the physical region and in the range of the 
lowest existing data) to fit the LECs. However, chiral perturbation theory is 
expected to yield the most reliable predictions for $s$ (the center--of--mass energy
squared)  and $t$ (the squared invariant four--momentum transfer) lying 
{\it inside} the Mandelstam triangle (depicted in fig.\ref{fig:mandel}), 
for essentially two reasons. First,
in this region the scattering amplitude is purely real and it is well known
that at a given order in the chiral expansion, the real part is in general more
precisely determined than the corresponding imaginary part (since the latter only
starts at one loop order). Second, in the interior of  the Mandelstam triangle
the kinematical variables $t$ and $(s-u)/4m$ take their smallest 
values.\footnote{Throughout, we denote by $m=938.27\,$MeV and $M=139.57\,$MeV 
the nucleon and the pion mass, respectively, and the Mandelstam variables are 
subject to the constraint $s+t+u=2m^2+2M^2$. Furthermore, we work in the isospin limit
$m_u=m_d$ and neglect all virtual photon effects. The exception to this
are the isospin violation effects contributing to the particle masses, i.e. kinematical
effects.} As this region is unphysical, there is no 
direct access by experimental data. By the use of dispersion relations this 
problem can be circumvented. This is done here. First, using data from phase
shift analysis we construct the pion--nucleon amplitude inside the Mandelstam 
triangle. We then use the chiral third order amplitude constructed in~\cite{FMS}
to determine the 8 LECs under consideration by a best fit as described
below and compare their numerical values with the ones obtained previously.
The ninth LEC is fixed by the value of the pion--nucleon coupling constant used
in our analysis.
We expect a more precise determination of the dimension two LECs since to the
order we are working, we are not sensitive to the $1/m$ corrections to the
dimension three LECs (which are known to be important in the determination
of the $c_i$, compare e.g. the values obtained in~\cite{bkmppn} with
the ones of refs.\cite{BKMpin,moj,FMS}.). 
We remark that H\"ohler~\cite{GH} has stressed that one should 
also compare directly
the analytical form of the pion--nucleon amplitude obtained by different
means. We refer to his work for a detailed comparison between the dispersive
and the chiral representation. Also, it is known that in some small regions the
heavy baryon amplitude converges slowly. This can be traced back to the
fact that the strict heavy baryon limit tends to modify the analytical
structure of the $\pi N$ amplitude. 
These effects can be dealt with by subtracting from the amplitudes the
full Born terms, since the latter generate the singularities
(such an effect also appears in the discussion
of the nucleon electromagnetic form factors,
see~e.g.~\cite{GSS,BKMff,ulfhugs}). It is important to stress that
this subtraction procedure is not arbitrary
since the subthreshold expansion of the $\pi N$ invariant amplitudes
is usually formulated by subtracting the Born terms to avoid their
rapid variations in the appropriate kinematical variables.
Another way of circumventing this
problem is to stick to a relativistic formulation of the matter fields,
as recently proposed in ref.\cite{BL} since in that way all strictures
from analyticity are automatically fulfilled.  Here, however, we are concerned
with the comparison of the $\pi N$ amplitude obtained in 
heavy baryon chiral perturbation theory (HBCHPT) since in
this framework the most precise predictions for the threshold parameters
have been obtained and we wish to explore the consistency of these
calculations at the order they have been performed. In contrast to previous
investigations, we confine ourselves to the inside of the Mandelstam triangle
for the reasons mentioned above.

\medskip
\noindent The manuscript is organized as follows. Some formalism pertaining
to elastic pion--nucleon scattering pertinent to our investigation is given
in sec.~\ref{sec:form}. In sec.~\ref{sec:cons} we construct the invariant
amplitudes inside the Mandelstam triangle by use of dispersion relations.
The chiral perturbation theory amplitudes are fitted to these dispersive 
amplitudes in sec.~\ref{sec:fits}. Further results on subthreshold parameters
and the $\sigma$--term are discussed in sec.~\ref{sec:res}. The summary and
conclusions follow in sec.~\ref{sec:summ}.

\section{Formal aspects of elastic pion--nucleon scattering}
\label{sec:form}

Consider elastic pion--nucleon scattering, 
\beq
\pi^a (q_1) + N(p_1)\to \pi^b (q_2) + N(p_2)~,
\eeq
with `$a,b$' (cartesian) pion isospin indices and $q_i$, $p_i$ the four--momenta
of the pions and the nucleons, respectively. The
scattering amplitude is usually decomposed in terms of the four invariant
functions $A^\pm (s,t)$ and $B^\pm (s,t)$ (where the superscript `$\pm$'
refers to the isoscalar/isovector part),
\beqa
T^{ba}_{\pi N} (s,t) &=& T_{\pi N}^+ (s,t) \, \delta^{ba} + T_{\pi N}^- (s,t) \,
\frac{1}{2} [\tau^b, \tau^a]~, \nonumber\\
T^\pm_{\pi N} (s,t)  &=& 
\bar{u}(p_2,\lambda_2) \, \biggl[ A^\pm (s,t) + \frac{1}{2}(q_2 +q_1)^\mu \gamma_\mu \,
B^\pm (s,t)\biggr] \, u(p_1,\lambda_1)~,
\eeqa
with $\sqrt{s}$ the cms energy and $t=(q_1-q_2)^2$ the invariant momentum 
transfer squared. The $\lambda_i$ ($i=1,2)$ denote the helicities of the
incoming/outgoing nucleon.
In what follows, we also need the linear combinations related to the
physical channels $\pi^+ p \to \pi^+ p$ and  $\pi^- p \to \pi^- p$.
These are related to the isospin amplitudes by
\beq\label{X}
X^\pm (s,t) = \frac{1}{2} \biggl( X_-(s,t) \pm X_+(s,t)\biggr)~,
\quad X=\{ A,B\}~.
\eeq
The invariant amplitudes fulfill the crossing relations
\beqa
   A^{+}(s,t,u) & = & A^{+}(u,t,s)~,\nonumber \\
   B^{-}(s,t,u) & = & B^{-}(u,t,s)~,\nonumber \\
   A^{-}(s,t,u) & = & -A^{-}(u,t,s)~,\nonumber \\
   B^{+}(s,t,u) & = & -B^{+}(u,t,s)~.
\eeqa
For heavy baryon chiral perturbation theory, it is more natural to work
directly with the non--spin--flip and spin--flip amplitudes $g^\pm (\omega,t)$
and $h^\pm (\omega,t)$, respectively. For a discussion of this point, see~\cite{FMS}.
Then, the $\pi N$ scattering amplitude takes the form
\beqa 
T^{ba}_{\pi N} (\omega , t) &=& 
\biggl(\frac{E+m}{2m}\biggr) \, \biggl\lbrace \delta^{ba} 
\Big[ g^+(\omega,t)+ i \vec
\sigma \cdot(\vec{q}_2\times \vec{q}_1\,) \, h^+(\omega,t) \Big]
\nonumber \\ && \qquad\quad
+i \, \epsilon^{bac}
\tau^c \Big[ g^-(\omega,t)+ i \vec \sigma \cdot(\vec{q}_2 \times \vec{q}_1\,) \,
h^-(\omega,t) \Big] \biggr\rbrace 
\eeqa
with $\omega$ the pion cms energy, 
\beq\label{om}
\omega 
= \sqrt{\frac{(s-m^2+M^2)^2}{4 s}}~,
\eeq
$E  = ( \vec{q\,}^2 +m^2)^{1/2}$ the nucleon energy and the
pion momenta
\beq
\vec{q\,}_1^2 = \vec{q\,}_2^2 \equiv
\vec{q\,}^2 \equiv q^2 = {(s-M^2-m^2)^2 -4m^2M^2 \over 4s} \quad .
\eeq
The relations between the invariant amplitudes $A^{\pm},B^{\pm}$ and $g^{\pm},h^{\pm}$
is
\beqa
\label{eq:inv_flip1}
g^\pm (\omega,t) = {C_4 A^\pm (\omega,t) - C_2 B^\pm (\omega,t) 
\over C_1C_4 - C_2C_3}~,\\
\label{eq:inv_flip2}
h^\pm (\omega,t) = {-C_3 A^\pm (\omega,t) + C_1 B^\pm (\omega,t) 
\over C_1C_4 - C_2C_3}~,
\eeqa
with the coefficient functions $C_i$ $(i=1,\ldots,4$) given by
\begin{eqnarray}\label{Ci}
  C_1 & = & {\frac{{m^2} - {M^2} - m\,\omega  + 
     m\,{\sqrt{{m^2} - {M^2} + {{\omega }^2}}}}{2\,({m^2} - 
     {M^2})}}~, \nonumber \\
  C_2 & = & \frac{1}{4\,  \left( m^2 - M^2 \right)
     }\left\{4\,{m^2}\,{M^2} - 4\,{M^4} - {m^2}\,t + {M^2}\,t - 
 4\,{m^3}\,\omega\right.\nonumber  \\ &  & \left.+ m\,t\,\omega  + 
     4\,{M^2}\,{{\omega }^2} - 
     m\,\left(  t + 4\,m\,\omega -4\,{M^2}  \right) \,
      {\sqrt{{m^2} - {M^2} + {{\omega }^2}}}\right\}~, \nonumber\\
   C_3 & = & {\frac{1}{2\,\left( \omega  + 
       {\sqrt{{m^2} - {M^2} + {{\omega }^2}}} \right) }}~, \nonumber\\
   C_4 & = & {\frac{\left( m + {\sqrt{{m^2} - {M^2} + {{\omega }^2}}} \right)
       \,\left(  4\,m\,
        \left( m + {\sqrt{{m^2} - {M^2} + {{\omega }^2}}}
          \right) -t \right) }{4\,
     \left( {m^2} - {M^2} + m\,\omega  + {{\omega }^2} + 
       \left( m + \omega  \right) \,
        {\sqrt{{m^2} - {M^2} + {{\omega }^2}}} \right) }}~.
\end{eqnarray}
This form differs from the one in~\cite{FMS} in that it is
explicitly given in terms of $\omega$ and $t$ and also it is
more directly applicable for considering processes in regions other than the physical
$s$--channel domain. Note that these coefficient functions should not be expanded in
$1/m$ when one performs the chiral expansion. Finally, we will also 
use the crossing--symmetric
variable $\nu = (s-u)/ 4m$, the scattering angle in the
center--of--mass system in the $s$-channel, $\cos \theta =
1+t/2q^2$, and sometimes denote the cms energy by $W$, i.e. $s=W^2$.
We also need the subthreshold expansion of the invariant amplitudes
with the pseudovector Born terms subtracted (as indicated by the
``bar'')~\cite{hoeh}
\beq
\bar{X} = \sum_{m,n} x_{mn} \, \nu^{2m+k} \, t^n~,\quad X=\{A^\pm,B^\pm\}~,
\eeq
with $k=1 \,(0)$ if the function considered is odd (even) in $\nu$.
The Taylor--coefficients $x_{mn}$ are the so--called subthreshold parameters.
For a more detailed account of the pertinent kinematics, we refer to the
monograph~\cite{hoeh}.

\medskip

\section{Construction of the invariant amplitudes inside the
Mandelstam triangle using dispersion relations}
\label{sec:cons}

In the next step, we use dispersion relations to construct
the $\pi N$ amplitude inside the Mandelstam triangle. The Mandelstam
triangle is the interior of the region bounded by the three lines
$s=(M+m)^2$, $u=(M+m)^2$ and $t = 4M^2$ in the Mandelstam plane, see
fig. \ref{fig:mandel}. 
Dispersion relations for $\pi N$ scattering have been studied in great
detail in axiomatic field theory in the last thirty years~\cite{disprel,hoeh}. 
Using analyticity, unitarity, and crossing symmetry one obtains dispersive
relations for the scattering amplitudes of the form
\beqa
{\rm Re}~A^\pm (\nu,t) &=& {1 \over \pi} {\int \!\!\!\!\!\!-}_{\nu_{\rm thr}}^\infty 
\!\!\!\!\!{\rm Im}~A^\pm (\nu',t) \biggl\{ \frac{1}{\nu'-\nu} \pm
\frac{1}{\nu'+\nu}\biggr\}\, d\nu'~, \\ \label{BDR}
{\rm Re}~B^\pm (\nu,t) &=& {1 \over \pi} {\int \!\!\!\!\!\!-}_{\nu_{\rm thr}}^\infty 
\!\!\!\!\!{\rm Im}~B^\pm (\nu',t) \biggl\{ \frac{1}{\nu'-\nu} \mp
\frac{1}{\nu'+\nu}\biggr\}\, d\nu' 
+ \frac{g^2}{2 m} \left(\frac{1}{\nu_B - \nu} \mp
  \frac{1}{\nu_B+\nu}\right)~,
\nonumber \\ &&
\eeqa
with $\nu_{\rm thr} = M + t/(4 m)$ and $\nu_B = (t-2 M^2)/(4 m)$. Note
that the nucleon Born terms contribute only  to $B^\pm (\nu,t)$. Here
\beq\label{fval}
f^2 = \frac{g^2}{4\pi}\left( \frac{M}{2m}\right)^2 \simeq 0.079
\eeq
is the pion--nucleon coupling constant squared. There is some debate about its
actual value, some analysis favoring a somewhat smaller value, $f^2 \simeq
0.075$, as discussed below.
However, to be consistent with refs.\cite{BKMpin,moj,FMS}, which mostly use the 
Karlsruhe phase shifts and therefore the value given in eq.(\ref{fval}),
we also use the large value for $f^2$. Note that in addition we consider an analysis
favouring a smaller value for $f^2$ later on.  
We mention that it seems clear that the imaginary
part of the amplitudes $A^-, B^\pm$ fall off sufficiently fast so that no 
subtraction is needed. This might also be true for $A^+$, see e.g.~\cite{HS}. 
However, to precisely determine the subthreshold parameters related to $A^+$, 
we perform a subtraction for the amplitude
$A^+$~\cite{hoeh},
\beq\label{Aps}
{\rm Re}~A^+ (\nu,t) = {\rm Re}~A^+ (\nu=0,t) +
 {2\nu^2 \over \pi} {\int \!\!\!\!\!\!-}_{\nu_{\rm thr}}^\infty 
\, {{\rm Im}~A^+ (\nu',t) \over \nu'(\nu'^2-\nu^2)}\, d\nu'~, 
\eeq
where the first term on the right--hand--side of eq.(\ref{Aps}) is the 
subtraction function. This will be discussed in more detail below.
Note also  that the influence of any high
energy contribution is damped further in the once subtracted dispersion relation
for $A^+ (\nu,t)$.

\medskip

\noindent
The absorptive parts of the amplitudes in the dispersive integrals are
broken up into two parts: a) the low energy part with laboratory momenta
$k_{\rm lab} \le 6\,$GeV
and b) the high energy part. The first  is constructed by partial waves
of the Karlsruhe group (KA84)~\cite{KA84}. We note that there
has been considerable criticism of these partial wave amplitudes
recently (see e.g. the proceedings of MENU~97~\cite{menu97}), however,
at present no other analysis exist which consistently includes data
from threshold to the highest available energies (for example, the VPI
group uses the Karlsruhe analysis for $k_{\rm lab}$ above 2.1~GeV \cite{GH}).  
In our numerical analysis we work with the S-- to K--wave approximation
of the physical amplitudes $A_\pm (\nu,t)$ and $B_\pm (\nu,t)$, i.e.
(for clarity, we only display the S-- and P--wave contributions here.
The ellipsis stands for the contributions from the higher partial waves.)
\beqa
{A_+ (\nu,t) \over 4\pi} &=& {W+m\over E+m} \biggl\{f_{0+}^{3/2} (\nu) + 
3 \cos \theta f_{1+}^{3/2}(\nu)\biggr\} - {W-m\over E-m} \biggl\{
f_{1-}^{3/2}(\nu)-f_{1+}^{3/2}(\nu)\biggr\}~+\ldots, \nonumber \\ &&\\
{A_- (\nu,t) \over 4\pi} &=& {W+m\over 3(E+m)} \biggl\{f_{0+}^{3/2} (\nu) +
2f_{0+}^{1/2} (\nu) -3\cos\theta \bigl[f_{1+}^{3/2} (\nu) -2f_{1+}^{1/2} (\nu) 
\bigr]\biggr\} \nonumber \\
&-& {W-m\over 3(E-m)} \biggl\{f_{1-}^{3/2} (\nu) +  2f_{1-}^{1/2} (\nu)
- f_{1+}^{3/2} (\nu) -2f_{1+}^{1/2} (\nu) \biggr\}~+\ldots,   \\
{B_+ (\nu,t) \over 4\pi} &=& {1\over E+m} f_{0+}^{3/2} (\nu) + \biggl\{
{3 \cos \theta\over E+m} - {1\over E-m}\biggr\} f_{1+}^{3/2}(\nu) + {1\over E-m}
f_{1-}^{3/2}(\nu)~+\ldots, \\
{B_- (\nu,t) \over 4\pi} &=& {1\over 3(E+m)} \biggl\{f_{0+}^{3/2} (\nu) +
2f_{0+}^{1/2} (\nu) +3\cos\theta \bigl[f_{1+}^{3/2} (\nu) +2f_{1+}^{1/2} (\nu) 
\bigr]\biggr\} \nonumber \\
&+& {1\over 3(E-m)} \biggl\{f_{1-}^{3/2} (\nu) +  2f_{1-}^{1/2} (\nu)
- f_{1+}^{3/2} (\nu) -2f_{1+}^{1/2} (\nu) \biggr\}~+\ldots~,
\eeqa 
yielding the desired amplitudes $A^\pm, B^\pm$ by eq.(\ref{X}). 
To construct e.g. the amplitude $B_+ (\nu,t)$, we need the partial
waves with the appropriate quantum numbers from S31 up to 
K315  (in the usual notation $l_{2I,2j}$). 
The pertinent partial wave amplitudes $f_{l\pm}^I (\nu)$ for total
isospin $I=1/2,3/2$, pion--nucleon angular momentum $l$ and total
angular momentum $j = l\pm 1/2$ are given
in terms of the phase shifts $\delta_{l\pm}^I (\nu)$ and inelasticities
 $\eta_{l\pm}^I (\nu)$ via
\beq
f_{l\pm}^I (\nu) = {1\over 2iq} \bigl\{ \eta_{l\pm}^I (\nu) \exp 
\delta_{l\pm}^I (\nu) -1 \bigr\}~.
\eeq

\begin{figure}[htb]
\psfrag{xxx}[Bl][][0.8]{\hspace{-3mm}$\mbox{Re }A^{+}(\nu,t=0)$}
\psfrag{yyy}[][][0.8]{$\mbox{Im }A^{+}(\nu,t=0)$}
\begin{center}
\epsfig{file=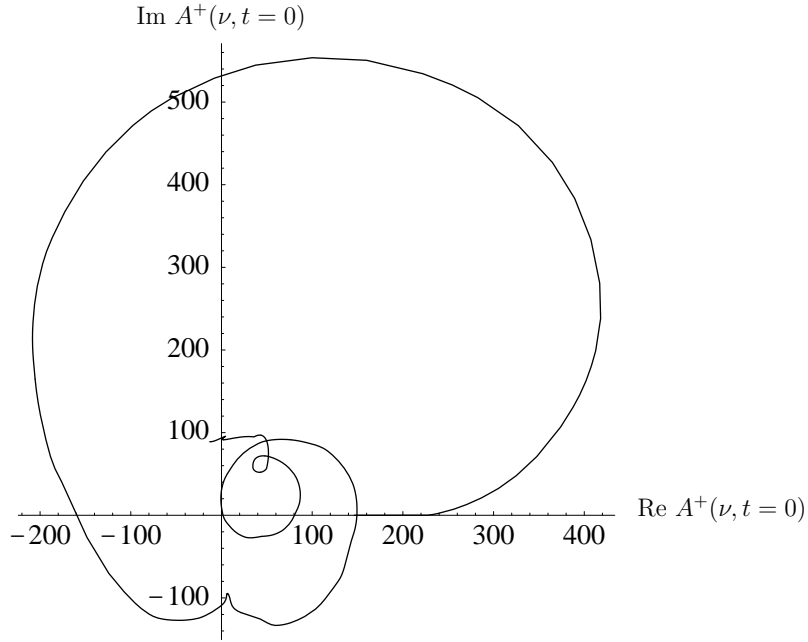,height=9.cm}
\caption{Argand plot of $A^+(\nu,t=0)$ for $k_{\rm lab}\leq 6\,$GeV. All units are
appropriate powers of the pion mass.}
\label{fig:argA}
\end{center}
\end{figure}

\begin{figure}[htb]
\psfrag{xxx}[Bl][][0.8]{\hspace{-3mm}$\omega\ \mbox{Re }B^{-}(\nu,t=0)$}
\psfrag{yyy}[][][0.8]{$\omega\ \mbox{Im }B^{-}(\nu,t=0)$}
\begin{center}
\epsfig{file=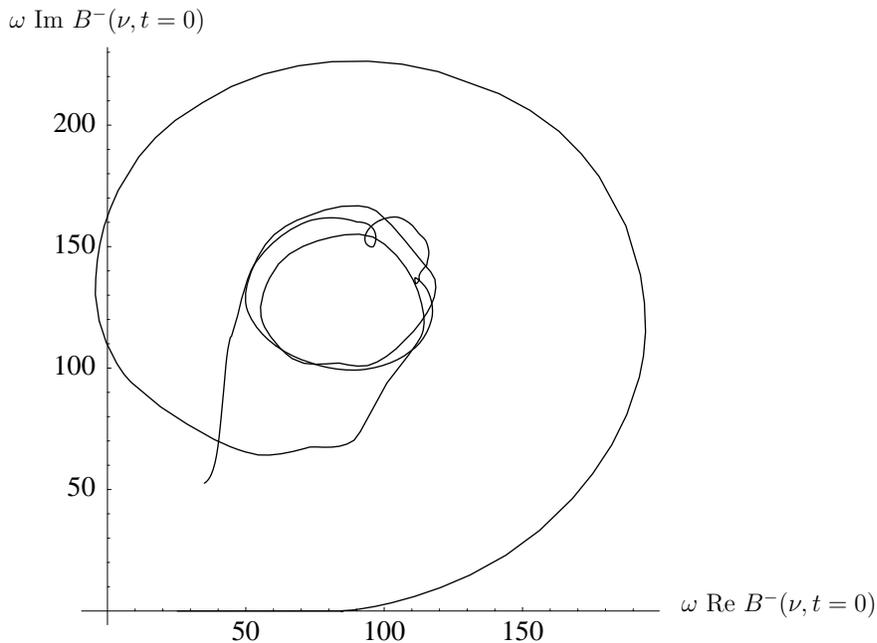,height=9.cm}
\caption{Argand plot of $\omega \, B^-(\nu,t=0)$ for $k_{\rm lab}\leq 6\,$GeV. All units
are appropriate powers of the pion mass.}
\label{fig:argB}
\end{center}
\end{figure}
\noindent
The corresponding Argand diagrams for $A^+$ and $\omega B^-$ in the
low energy region are shown in figs.~2 and 3, respectively. These
reproduce the ones 
given by H\"ohler and collaborators with sufficient accuracy for our later purposes,
compare with the corresponding figures in ref.\cite{physdata}.
For the high energy part of the amplitudes $A^-$ and $B^-$ 
we assume a reggeized $\rho$--meson
exchange in the $t$--channel of the charge exchange process 
$\pi ^{-} p \to \pi^{0} n$ to be the sole component of the amplitude~\cite{hoeh},
e.g.
\beq
{\rm Im}~B_R^- (\omega, t) = D_0 \, \alpha(t) \{ \alpha(t) +1 \} \biggl(
\frac{\omega}{\omega_0} \biggr)^{\alpha(t)-1}~,
\eeq
with $D_0 = 38.9\,$mb, $\omega_0 = 1\,$GeV and the Regge slope is parameterized via
$\alpha(t) = 0.56 + 1.08 \cdot t$~\cite{spear}.

\begin{figure}[h]
\psfrag{xxx}[Br][][0.9]{$t$}
\psfrag{yyy}[][][0.9]{$A^{+}(\nu=0,t)$}
\begin{center}
\epsfig{file=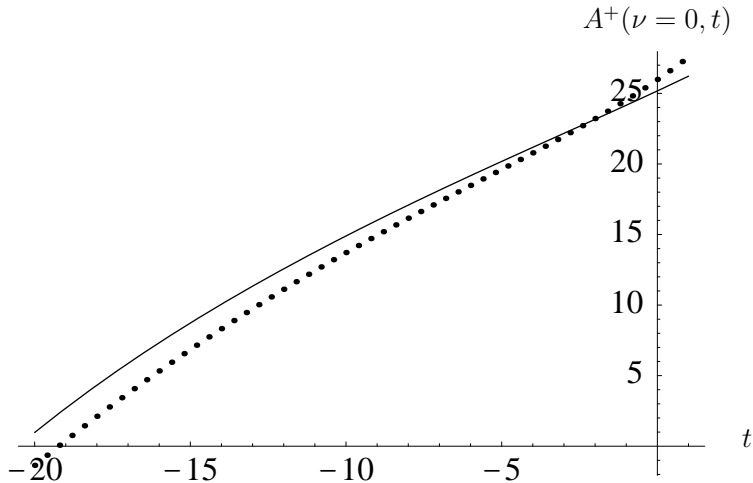,height=6.5cm}
\caption{The subtraction function $A^+(\nu=0,t)$. The solid (dotted) line
refers to the KA84 (VPI SP99) partial--wave analysis.}
\label{fig:subtr}
\end{center}
\end{figure}

\noindent
In the present work we will make use of the amplitudes along the lines
$\nu =0$ and $t=2M^2/3$ (and also $t=0$) to map out the interior of the 
Mandelstam triangle (cf. fig.~1). The choice of this particular value
of $t$ is justified in the next section.
Consequently, $B^- (0,t)$ and the subtraction function $A^+(0,t)$ 
will be of special interest to us. We remark that $A^- (0,t)$ and $B^+
(0,t)$ vanish because
of crossing symmetry. $A^+(0,t)$ is reconstructed by approximating the real and
imaginary part of $A^+$ in the physical region of the $s$--channel
 by the available partial waves. Inverting eq.(\ref{Aps})
then yields $A^+ (0,t)$.\footnote{Note that inside the Mandelstam triangle the
amplitude is real.}  The result is shown in fig.~\ref{fig:subtr}. 
We remark that this amplitude fulfills the Adler consistency condition
\beq\label{Adler}
A^+(\nu=0,t=M^2,q_1^2=0,q_2^2=M^2) \simeq \frac{g^2}{m}
\eeq
within one percent. Since physical pions do not have vanishing four--momentum, one
expects a deviation from this relation of the order of $M^2/(4\pi F)^2 \simeq
1.5\%$.
$B^- (0,t)$ is obtained from eq.(\ref{BDR}). 
The $\nu$--dependence of the amplitudes is evaluated in a
similar fashion by keeping $t$ fixed at the value given above. 

\medskip
\noindent So far, we have used the KA84 phase shift analysis. We are
aware that some of the data entering that analysis are now considered
inconsistent. Consequently, we will also use the latest version of
the VPI phase shift analysis, called SP99~\cite{VPI}, for our calculations.
We note that the VPI group uses dispersion relations only for some
of the amplitudes and in a certain energy region, different from the
Karlsruhe method. 
Based on the VPI SP99 phases, we have constructed
the amplitudes $A^\pm, B^\pm$ as described before. 
We refrain from showing the corresponding
figures, but we remark that the Adler consistency condition is then
violated by 10 percent. The corresponding subtraction function is also
shown in fig.\ref{fig:subtr} by the dotted curve. Of course, one has to 
account for the smaller value of the pion--nucleon coupling
constant used by the VPI group, $f^2 \simeq 0.076$. The Adler
consistency relation is of
importance for the precise determination of the amplitudes when the
Born terms are subtracted. We believe that for making accurate statements
on small quantities like the pion--nucleon $\sigma$ term, it is mandatory
to fulfill such constraints as e.g. the Adler condition to the
expected accuracy as explained before.\footnote{We have been informed
  by Marcello Pavan that the
  VPI group is presently improving their representation of the
  subtraction function.}

\medskip

\section{Chiral perturbation theory amplitudes inside the Mandelstam
triangle}
\label{sec:fits}

We now turn to the comparison with the chiral
amplitude. For that, it is important to study the dependence of
the various invariant amplitudes on the pertinent LECs. We give here
the explicit expressions for the counterterm amplitudes taken from
ref.\cite{FMS} (in terms of $t$ and $\omega$) 
\beqa
g^+_{\rm ct} &=& \frac{c_2 \, \omega}{m F^2} \Bigl[ 4 \omega^2 - 4M^2
+t \Bigr]+\frac{1}{F^2}\left[ -4 c_1 M^2 + 2 c_2 \omega^2 + c_3 (2 M^2-t)\right]~, \\
g^-_{\rm ct} &=& \frac{c_4 \, \omega \, t}{2m F^2} 
+ \frac{2\omega}{F^2} (2M^2-t) \, (\bar{d}_1+\bar{d}_2 )
+ \frac{4\omega^3}{F^2} \, \bar{d}_3 +\frac{8 \omega\, M^2}{F^2} \, \bar{d}_5
\nonumber \\ &&
\qquad\qquad 
+ \,
\frac{g_A}{F^2}\frac{M^2}{\omega} (2\omega^2-2M^2+t)\,
\bar{d}_{18} \\
h^+_{\rm ct} &=& \frac{2 \omega}{F^2} \, (\bar{d}_{14} - \bar{d}_{15}
) + \frac{g_A}{F^2} \frac{2 M^2}{\omega}\,\bar{d}_{18}~,\\
h^-_{\rm ct} &=& \frac{c_4 \, \omega}{m F^2} + \frac{c_4}{F^2}~,
\eeqa
to facilitate the later discussion. The $\nu$--dependence can be worked out
using the fact that $\omega$ and $\nu$ are related via
\beq
\nu = \frac{1}{4m} \, (2s+t -2M^2 -2m^2)~,
\eeq
and using eq.(\ref{om}). Note that most of the LECs appear 
only in one particular invariant amplitude, e.g. $c_1,c_2,c_3$ are fixed entirely from
varying $g^+ (\omega, t)$ inside the Mandelstam triangle. Only
$c_4$ and $\bar{d}_{18}$ appear in two different amplitudes.
In particular, we expect that most of the dimension three LECs cannot
be pinned down accurately, since they only appear in $g^-_{\rm ct}$
with small prefactors. This expectation is borne out by the results
given below. Furthermore, to this order in the chiral expansion we are
not sensitive to the $1/m$ corrections to the $\bar{d}_i$. Such terms are
known to play an important role in the precise determination of the
dimension two LECs, compare e.g. the values obtained in ref.\cite{bkmppn}
with the ones in refs.\cite{BKMpin,moj,FMS}.
It should be noted that the amplitudes given in~\cite{FMS} are valid for $\omega>M, t<0$.
Therefore, when working inside the Mandelstam triangle,
one has to ensure the correct analytic continuation to the unphysical region $\omega <M$
of the complex--valued one--loop
contributions of $g^\pm$ and $h^\pm$ by use of~\cite{BKMpin}
\beq
\sqrt{1-x^2} = -i \, \sqrt{x^2-1} ~, \quad
\arcsin x = \frac{\pi}{2} + i \ln \left( x + \sqrt{x^2-1} \right)~.
\eeq
A similar statement holds for the continuation to $t>0$.
We use the tree and one--loop amplitudes
explicitly given in~\cite{FMS} but refrain from spelling them out here.

\smallskip
\noindent
The chiral amplitudes to order $O(q^3)$ do not have
the correct analytic structure inside the Mandelstam triangle. This is
in part due to the Taylor expansion of the
lowest order amplitude in powers of $m^{-1}$. To avoid this problem,
it is necessary to neglect the tree contributions and
the counter terms $\sim g_A$. This holds in particular for the
tree graphs with insertions $\sim \bar{d}_{18}$. The latter allow one
to use the physical value of the pion--nucleon coupling constant
(instead of its value in the chiral limit) in the tree graphs, which
are responsible for the incorrect analytic behaviour.
This procedure is equivalent to working with the quantities
$\bar{A}^+\equiv A^+ - g^2/m$ and so on instead of $A^{\pm}, B^{\pm}$ in
eqs.(\ref{eq:inv_flip1}),(\ref{eq:inv_flip2}). This way, all the contributions due to
$\bar{d}_{18}$ drop out and the latter has to be calculated in a
different way, e.g. by the so--called Goldberger-Treiman discrepancy
(i.e the deviation from the Goldberger-Treiman relation):
\beq
g_{\pi N} = \frac{g_A m}{F_\pi} \left (1-\frac{2 M_\pi^2 \bar{d}_{18}}{g_A}\right )~.
\eeq
In the present work we compare the chiral amplitudes with their
dispersive counterparts in a small region inside the Mandelstam
triangle. More precisely, we concentrate on two regions. The first of these
is concentrated around the
point $\nu =0$ and $t=2M^2/3$, the so--called ``ideal point''. 
In the chiral limit, where the pions are massless, the ideal point
is given by $\nu = t =0$ and the constraint $s+t+u = 2m_0^2$ has to be
fulfilled, with $m_0$ the nucleon mass in the chiral limit. Since we
are considering physical nucleons and physical pions, the ideal point 
is shifted to  $t \simeq 2M^2/3$ because now $s+t+u = 2m^2 + 2M^2$. 
We consider two fits, which span the following regions within 
the Mandelstam triangle 

\beqa\label{fitr}
{\rm Fit~1:} &&  {\rm around}~\nu =0,~t=0~\no\\
&& {\rm with}~\{-0.5~M^2 \le t \le 0.5~M^2\,
,45.7M^2 \leq s \leq 46.7M^2\}~,\\
{\rm Fit~2:} &&  {\rm around}~\nu =0,~t=2M^2/3~\no\\
&& {\rm with}~\{-0.3~M^2 \le t \le 0.7~M^2 \,
,45.35M^2 \leq s \leq 46.35\}~,
\eeqa
This means that for both fits we cover the range of one pion
mass squared in the two independent Mandelstam variables. A larger range
cannot be accounted for by the third order chiral amplitudes.
We have convinced ourselves that decreasing these
ranges by a factor of two does not change our results. 
Note that for the fits along the line $\nu =0$, a simplification
arises. Due to
crossing symmetry and eqs.(\ref{eq:inv_flip1}),(\ref{eq:inv_flip2}),
each of the amplitudes $\bar{g}^{\pm},
\bar{h}^{\pm}$ depends on only one of the invariant amplitudes
$\bar{A}^{\pm}, \bar{B}^{\pm}$. E.g. the non-spin-flip amplitude
$\bar{g}^{+}(\nu=0,t)$ is related to the modified subtraction function
$\bar{A}^{+}(\nu=0,t)$ via
\be
   \bar{g}^{+}(0,t) & = & \frac{C_4}{C_1 C_4 - C_2 C_3}
   \bar{A}^{+}(0,t)~.
\ee

\medskip
\noindent
A fit  to each of the dispersive amplitudes $\bar{g}^{\pm}(\nu,t),
\bar{h}^{\pm}(\nu,t)$ at 21 equidistantly separated points in the two 
directions of fixed $\nu$ and fixed $t$ in the regions given in
eq.(\ref{fitr})  yields the following values for the LECs (we give
here the averaged values obtained from fit~1 and fit~2 together, the
individual results are displayed and discussed below):
\be
   \label{eq:res_fit}
   c_1 & = & (-0.81 \pm 0.12)~ \mbox{GeV}^{-1}~, \quad 
   c_2   =  (8.43 \pm  56.9)~ \mbox{GeV}^{-1}~,\nonumber\\
   c_3 & = & (-4.70 \pm 1.16)~ \mbox{GeV}^{-1}~, \quad
   c_4   =  (3.40 \pm 0.04)~ \mbox{GeV}^{-1}~. 
\ee
The errors given are the the ones from the two fits added in
quadrature. The errors for each fit have been determined from the
square root of the diagonal elements of the error matrix. 
We note that there is a sizeable variation in the precision with 
which the dimension two LECs are determined. 
It is expected that the isovector amplitudes
are more accurately predicted at third order by chiral
perturbation theory since the isoscalar amplitudes vanish to leading
order in the chiral expansion.\footnote{Note that this does not
   neccessarily imply a similar pattern for the accuracy with which
   the corresponding counterterm contributions can be pinned
   down. This argument can also be invalidated in case of zero
   crossings of certain amplitudes.}
That is also the reason why the 
LEC $c_4$ is more precisely pinned down than the LECs $c_{1,2,3}$.
In table~\ref{tab:ci} we compare the values of the dimension two LECs for fit~1 
and fit~2 separately
with previous determinations. 
\begin{table}[h]
\begin{center}
\begin{tabular}{|c|c|c|c|c|c|}
\hline
$c_i$ & Fit~1 & Fit~2 & Ref.\cite{BKMpin} & Ref.\cite{moj} & Ref.\cite{FMS} (Fit~1)\\ 
\hline
1 & $-0.81 \pm 0.15$ & $-0.80 \pm 0.07$ &
$-0.93 \pm 0.10$ & $-0.94 \pm 0.06$ & $-1.23 \pm 0.16$ \\
2 & $9.35 \pm 66.7$ & $7.52\pm 45.0$ &
$3.34 \pm 0.20$ & $3.20 \pm 0.10$ & $3.28 \pm 0.23$ \\
3 & $-4.69 \pm 1.34$ & $-4.70 \pm 0.95$ &
$-5.29 \pm 0.25$ & $-5.40\pm 0.06$ & $-5.94 \pm 0.09$ \\
4 & $3.40 \pm 0.04$ & $3.40 \pm 0.04$ &
$3.63 \pm 0.10$ & $3.47 \pm 0.06$ & $3.47 \pm 0.05$ \\
\hline
\end{tabular}
\caption{The dimension two LECs $c_{1,2,3,4}$ in GeV$^{-1}$ from our determination
based on the KA84 phase shifts compared to previous ones using also 
information from $\pi N$ scattering in the physical region and 
the $\sigma$--term.\label{tab:ci}}
\end{center}
\end{table}
\noindent
In ref.\cite{BKMpin}, (sub)threshold
parameters and the $\sigma$--term (i.e. quantities free of dimension
three LECs) were used, while in ref.\cite{moj} the $c_i$ and dimension
three LECs were determined from a fit to threshold parameters
of the Karlsruhe--Helsinki group~\cite{kopi}. Finally, the
determination in ref.\cite{FMS} (fit~1 therein) used only the available low--energy
KA84 phase shifts for pion laboratory momenta in the range from
40 to 97~MeV.
The results for $c_1$, $c_3$ and $c_4$ are very consistent for all the different
determinations.\footnote{Note that we come back to the different value for $c_1$ 
obtained here and from scattering data alone~\cite{FMS} in section~\ref{sec:res}.}
 The exception is our value for $c_2$, which is
essentially an undetermined quantity. There reason for that is that in
the amplitude $g^+$, which is small, the LECs $c_1$ and $c_3$ are
weighted with factors of $M^2$, whereas the term $\sim c_2$ is
proportional to $\omega^2$ (to leading order in the $1/m$)
and this quantity is suppressed by a factor
of 10 compared to $M^2$ in the region around the center of the
Mandelstam triangle. 
As a consequence, we can not make a precise prediction for the combination
$c^+ = c_2+c_3 -2c_1$, which can also be deduced from pion scattering off
deuterium~\cite{bblm}.  One could, however, turn the argument around and
use the value of $c^+$ determined in ref.\cite{bblm}, $c^+ 
= -0.09\pm 0.37\,$GeV$^{-1}$. Neglecting the uncertainties in $c_1$ and $c_3$,
this leads to 
\beq
c_2   =  (2.99 \pm 0.37)~ \mbox{GeV}^{-1}~,
\eeq
which is consistent with the other determinations compiled in table~\ref{tab:ci}.
It is also possible to pin down the LECs $c_{1,3,4}$ from the
long--range part of the proton--proton interaction based on the
chiral two--pion exchange potential. This has been done
in ref.\cite{nij} and leads to results consistent with the ones found
here and in previous works. The  values obtained in ref.\cite{nij} are
$c_1 = -0.76(7)$, $c_3 = -5.08(28)$ and $c_4=4.70(70)$ (all in GeV$^{-1}$).
Finally, we remark that the errors we quote (compare
eq.(\ref{eq:res_fit}) and table~\ref{tab:ci}) must
be handled with care: to our knowledge, there is no error analysis
available for the KA84 phase shifts and elasticities. Thus, it is
impossible to give reliable errors for the dispersive amplitudes
inside the Mandelstam triangle. In the present work we assume an
error of 10 \% for the amplitudes of interest. In this sense our
errors are arbitrary. Stated differently, 
using other errors does not change the central values
but the uncertainties deduced from the corresponding error matrix.

\medskip
\noindent
We now turn to the  dimension three LECs. We have found that if we fit
along lines of fixed $\nu$ or fixed $t$, they can be pinned down with
good precision. However, the so determined values are not mutually consistent.
The reason for this lies in the fact that the third order amplitude employed
here is not sufficiently precise to describe the small contributions from
${\cal L}_{\pi N}^{(3)}$ accurately. In table~\ref{tab:di} we compare our
results for the $\bar{d}_i$. Only the combination $\bar{d}_{14}-\bar{d}_{15}$ 
comes out consistent for the determinations within the regions given by
fits~1 and~2, respectively. Note, however, that the overall description
of the amplitude $h^+$ is fairly poor, so that this determination of
$\bar{d}_{14}-\bar{d}_{15}$ has to be taken with some caution 
(even though the uncertainty
in this LEC from the fits is tiny). It appears that
the $1/m^2$ corrections with fixed coefficients and the $c_i/m$ terms are
more important than the genuinely new operators of dimension
three.\footnote{A similar result for another set of dimension three
  LECs was recently reported in ref.\cite{fbm}.}
Also shown in table~\ref{tab:di} are results from previous determinations
determined by fitting to threshold parameters and phase shifts. It would be valuable
to have additional information on these LECs, say from a study of elastic
pion--nucleon scattering to fourth order in the chiral expansion  or
from the reaction $\pi N \to \pi \pi N$.
\begin{table}[h]
\begin{center}
\begin{tabular}{|c|c|c|c|c|}
\hline
$\bar{d}_i$ & Fit~1 & Fit~2  & Ref.\cite{moj} & Ref.\cite{FMS} (Fit~1)\\ \hline
1+2 & $3.33 \pm 0.01$ & $-1.91 \pm 0.14$ & $2.4 \pm 0.3$ & $3.06 \pm 0.21$  \\
3 &   $-152.3 \pm 0.24$ & $98.16 \pm 6.90$ & $-2.8 \pm 0.6$ & $-3.27 \pm 0.73$  \\
5 &   $-0.11 \pm 0.01$ & $-0.20 \pm 0.01$ & $1.4 \pm 0.3$ & $0.45\pm 0.42$  \\
14-15 & $0.96 \pm 0.00$ & $0.93 \pm 0.00$ & $-6.1 \pm 0.6$ & $-5.65 \pm 0.41$  \\
\hline
\end{tabular}
\caption{The dimension three LECs $\bar{d}_{i}$ in GeV$^{-2}$ (and combinations
thereof) from our determination
compared to previous ones using information from $\pi N$
scattering in the physical region.\label{tab:di}}
\end{center}
\end{table}

\medskip
\noindent
Using now the SP99 phases as input to construct the amplitudes and redo the fit, 
we find (taking into account the different value for the
$\pi N$ coupling constant used by the VPI~group)
\beq\label{VPI:fit}
\begin{array}{lcccr}
c_1 &=& (-3.00\ldots -2.96)~{\rm GeV}^{-1} &\quad& [(-1.53\pm 0.18)~{\rm GeV}^{-1}]~,\\
c_2 &=& (-9.51\ldots 7.22)~{\rm GeV}^{-1} &\quad&  [( 3.22\pm 0.25)~{\rm GeV}^{-1}]~,\\
c_3 &=& (-6.63\ldots -6.20)~{\rm GeV}^{-1} &\quad& [(-6.20\pm 0.09)~{\rm GeV}^{-1}]~,\\
c_4 &=& ( 3.39\ldots 3.40)~{\rm GeV}^{-1} &\quad& [( 3.51\pm 0.04)~{\rm GeV}^{-1}]~,\\
\bar{d}_{14} + \bar{d}_{15}
&=& ( 0.92 \ldots 0.93)~{\rm GeV}^{-2} &\quad& [(-1.53\pm 0.18)~{\rm GeV}^{-2}]~,\\
\end{array}
\eeq
where we have given the range based on fits~1,2 as described above.
The numbers in the square brackets are the results of fit~3 of ref.\cite{FMS}
based on the VPI SP98 solution. The uncertainties of the so determined LECs
are very similar to the ones obtained based on the KA84 partial waves. 
Again, we note that $c_2$ cannot be determined.
Notice in particular the very large value for $c_1$,
which will be commented on below. The value for $c_4$ is in perfect
agreement with the one found before, i.e.  the amplitude $h^-$ is the
same for the KA84 and VPI SP99 analysis. 
For $\bar{d}_1 + \bar{d}_2$, $\bar{d}_3$ and
$\bar{d}_5$ we encounter the same problems as discussed before and therefore
refrain from giving the corresponding numbers. We note, however, that the
values obtained for these LECs are very similar to the ones obtained for
the Karlsruhe phase shifts, compare table~\ref{tab:di}. 
For the reasons discussed in sec.\ref{sec:cons},
we consider the determination based on the KA84 phases more consistent (despite
the fact that a part of the data set used in the Karlsruhe analysis is now outdated).
As noted before, the dimension three LECs can mostly not be determined precisely.
For achieving a better precision, one has to go to fourth order and fit in a larger
region inside the Mandelstam triangle.

\section{Results and discussion}
\label{sec:res}

In HBCHPT the quantities of interest are expressed
in terms of the LECs. Explicit expressions for e.g. the subthreshold
parameters and the $\sigma$-term $\sigma(0)$ are given in \cite{BKMpin,FMS}.
For convenience, we collect some of these expressions in the appendix.
We concentrate here
on a subset of quantities namely these which a) were found to be most 
problematic in the analysis using mostly data from the physical
region and b) do only involve $\bar{d}_{14}-\bar{d}_{15}$ from the dimension
three LECs.\footnote{Luckily, these two conditions are mutually consistent.}
The data set eq.(\ref{eq:res_fit}) yields e.g. the following predictions for 
some of the low subthreshold coefficients 
(for the reasons mentioned above, we do not give any uncertainties):
\beq
a^{+}_{00} = -1.32~ M^{-1}~,\quad a^{+}_{10} = 4.49~ M^{-3}~,\quad
a^{+}_{01} =  0.97~ M^{-3}~,\quad b^{-}_{00} = 9.99~ M^{-2}~.
\eeq
The subthreshold parameters are in good agreement with the Karlsruhe analysis,
$a^{+}_{00} = -1.46\pm 0.10$,  $a^{+}_{10} = 4.66$\footnote{Note that
  the Karlsruhe group does not give an uncertainty for this quantity.},
$a^{+}_{01} = 1.14\pm 0.02$ and $b^{-}_{00} = 10.36 \pm 0.10$, in appropriate units of
the inverse pion mass.  Of particular interest is the result for
$a^+_{10}$, which came out consistently too large in the chiral analysis based on the
phase shifts, cf. table~E.1 in ref.\cite{FMS}. Furthermore, the result
for $b^{-}_{00}$, which has
not been given before in the literature, is  in good agreement with the
Karlsruhe analysis. While these results are very promising, one still
has to check their stability by performing a fourth order calculation.

\smallskip
\noindent
Most striking, however, is the result for the pion--nucleon $\sigma$--term, 
\beq\label{valsig}
\sigma(0) = 40~ \mbox{MeV}~,
\eeq
which agrees nicely with the
dispersion theoretical analysis of ref.\cite{GLS}, $\sigma (0) = 
(44\pm 8 \pm 7)\,$MeV, which is also based on the Karlsruhe phase shifts. On the
other hand, the third order chiral perturbation theory analysis based on the 
phase shifts lead to much larger values of the sigma--term, $\sigma (0) = 59\,$MeV 
in ref.\cite{moj} and $\sigma (0) = 70\,$MeV in ref.\cite{FMS} for the corresponding 
central values of the LEC $c_1$. This lends further credit to the statement that the 
chiral predictions for $\pi N$ scattering are most reliable inside the Mandelstam 
triangle since there the pertinent kinematical variables take their smallest possible 
values.  For the reasons mentioned above, we find it difficult to determine
a theoretical uncertainty to this number. We estimate the theoretical uncertainty 
(at this order) to be of the same size than in the analysis of ref.\cite{GLS}. Using the
formula derived in refs.\cite{jg,bm} 
\beq\label{y}
\sigma (0) = \frac{(36 \pm 7)~{\rm MeV}}{1-y}~, \quad
y = \frac{2\langle p| \bar{s}s|p\rangle}{\langle p| \bar{u}u +\bar{d}d |p\rangle}~,
\eeq
we get for the strangeness content $y = 0.10 \pm 0.17$, somewhat smaller
but compatible with the result of ref.\cite{GLS}.
If we insert the much larger value for $c_1$ obtained in the analysis based on
the VPI phase shifts, we obtain naturally a much larger $\sigma$--term of about
209~MeV. This value is even larger than the one found in ref.\cite{FMS}, where the
threshold $\pi N$ phases from the VPI SP98 solution were analyzed. As we noted 
before, this particular LEC (together with $c_3$) is sensitive to the subtraction
function $A^+ (0,t)$. We also stated before that the subtraction function deduced
from the SP99 solution leads to an unacceptably large deviation from the Adler
consistency condition. Consequently, the very small isoscalar amplitude might
not be well represented.  Inserting such a large  value for $c_1$ into eq.(\ref{y}), 
one  obtains a huge strangeness content,
$y = 0.83 \pm 0.04$, which would mean that a large fraction of the nucleon
mass is due to strange quark pairs. Such a picture of the nucleon is not tenable.
However, one might argue that higher orders not taken into account in eq.(\ref{y})
might drastically alter the number $(36\pm 7)\,$MeV in the numerator of eq.(\ref{y})
which is crucial in the link between $\sigma (0)$ and the strangeness content. 
While such a scenario is improbable, it can only be ruled out be a complete 
two--loop calculation of the scalar sector of baryon CHPT.  Clearly,
a full dispersive analysis using  the modern data set is needed to further
clarify this issue. We also mention that such a large $\sigma$--term
would be at odds with all other information one has on  nucleon matrix elements
of the various strangeness operators like e.g. $\bar{s}\, \gamma_\mu\, s$ or 
$\bar{s} \,\gamma_\mu  \gamma_5\,s $. The contribution of such matrix
elements to observables like e.g. the proton magnetic moment
seems to be fairly small, as indicated by recent
measurements on parity--violation in electron scattering~\cite{sample,happex}.
What also should be done is to supply more stringent theoretical 
uncertainties. For that,
one can not use the partial--wave analyses since these do not supply any error.
Consequently, one has to reanalyze the pion--nucleon scattering data. This
together with a more detailed discussion of the other amplitudes and subthreshold
parameters will be given elsewhere~\cite{BM2}.

\section{Summary and conclusions}
\label{sec:summ}

In this paper, we have considered pion--nucleon scattering inside the
Mandelstam plane from the point of dispersion and chiral perturbation
theory. Dispersion relations are based on general principles like
unitarity, crossing and analyticity and allow one to reconstruct
the invariant amplitudes  from the data, in particular also inside
the Mandelstam triangle. We have performed such a calculation here
based on the Karlsruhe partial--wave analysis as well as
on the SP99 solution of the VPI group. On the other hand,
chiral perturbation theory is the effective field theory of the
Standard Model at low energies and can be used to investigate the
strictures from the spontaneous and explicit chiral symmetry
violation. It is also based  on general principles and admits a
systematic power counting in small momenta and quark
masses. Therefore, it should work best inside the Mandelstam plane.
That this is indeed the case has been demonstrated here for heavy baryon
chiral perturbation theory (after subtraction of the Born terms which
need a special treatment because of the analytic structure). 
We have determined the so--called low--energy constants
by a direct comparison of the amplitudes obtained from HBCHPT with the
dispersive ones. From that, we can work out physical quantities and
in particular, we find a pion--nucleon $\sigma$--term of $\sigma (0) =
40\,$MeV, consistent with a previous dispersive analysis of
the Karlsruhe data~\cite{GLS}. We have also found a much improved description
of the subthreshold parameters $a^+_{10}$ and $b_{00}^-$. 
The VPI SP99 partial wave analysis leads
to a much larger $\sigma$--term of about 200~MeV. 
This partial wave analysis is, however, less stringently constrained
by strictures from  analyticity, which we believe to be an essential ingredient
for a precise determination of small quantities like the pion--nucleon 
$\sigma$--term. As noted before, the Karlsruhe analysis contains some data which
are now considered inconsistent with the rest of the data set. Therefore,
a full scale dispersive analysis based on the Karlsruhe method using only
the accepted modern data set is called for.

\bigskip

\section*{Acknowledgements}

\noindent We are grateful to Gerhard H\"ohler, Marcello Pavan
 and Mikko Sainio for useful communications and Nadia Fettes for some
 clarifying comments.

\bigskip

\appendix
\def\theequation{\Alph{section}.\arabic{equation}}
\setcounter{equation}{0}
\section{Expressions for some observables}
Here, we give the explicit expressions for the subthreshold parameters
and the $\sigma$--term discussed in sect.~\ref{sec:res}. These read
\beqa
a_{00}^+ &=&  \frac{2M^2}{F^2} \biggl( c_3-2c_1 \biggr) +
\frac{g_A^2 M^3}{8\pi F^4} \biggl(g_A^2 + \frac{3}{8}\biggr) 
+ {\cal O}(M^4)~, \\
a_{10}^+ &=&  \frac{1}{8\pi^2 \,F^2} \biggl(16\pi^2 F^2 c_2 -32\pi^2
mF^2 (\bar{d}_{14}-\bar{d}_{15} ) + mg_A^4\biggr) \no\\
&& \qquad\qquad\quad - \frac{M}{8\pi 
  F^4} \biggl(\frac{5g_A^4}{4} +1 \biggr)+ {\cal O}(M^2)~,\\
a_{01}^+ &=& -\frac{c_3}{F^2} - \frac{g_A^2 M}{16\pi F^4} 
\biggl(g_A^2 + \frac{77}{48}\biggr) + {\cal O}(M^2)~, \\
b_{00}^- &=& \frac{1}{2F^2} +  \frac{2m c_4}{F^2} -
\frac{g_A^2mM}{16\pi F^4} (1+g_A^2) + {\cal O}(M^2)~, \\
\sigma(0) &=& -4c_1 M^2 - \frac{9g_A^2M^3}{64\pi F^2} 
+ {\cal O}(M^4)~.
\eeqa
Note that we have corrected for a typographical error that 
appeared in eq.(E.5) of ref.\cite{FMS}.

\bigskip\bigskip\bigskip


\end{document}